\documentclass[11pt,a4paper]{article}
\usepackage{epsf,amsbsy,amssymb,graphicx}

\newcommand{\gr}[1]{\boldsymbol{#1}}
\newcommand{\be}{\begin{equation}}
\newcommand{\ee}{\end{equation}}
\newcommand{\bea}{\begin{eqnarray}}
\newcommand{\eea}{\end{eqnarray}}
\newcommand{\ket}[1]{|#1\rangle}

\title{Minimum~decoherence~cat-like~states\\ in~Gaussian~noisy~channels}
\author{A. Serafini${}^1$, S. De Siena${}^1$, F. Illuminati${}^1$, and 
M. G. A. Paris${}^2$\vspace*{.3cm}\\
\footnotesize ${}^1$ Dipartimento di Fisica ``E. R. Caianiello'', 
Universit\`a di Salerno, INFM UdR Salerno,\\
\footnotesize INFN Sez. Napoli, G. C. Salerno, 
Via S. Allende, 84081 Baronissi (SA), Italy \\ 
\footnotesize${}^2$I.S.I.S. ``A. Sorbelli",
via G. Matteotti 2, I-41026 Pavullo nel Frignano (MO), Italy}
\date{September 29, 2003}

\begin{document}
\maketitle
\begin{abstract}
We address the evolution of cat-like states in general Gaussian noisy 
channels, by considering superpositions of coherent and squeezed-coherent 
states coupled to an arbitrarily squeezed bath. The phase space dynamics is solved 
and decoherence is studied keeping track of the 
purity of the evolving state. The influence of the choice of the state and channel parameters 
on purity is discussed and optimal working regimes that minimize the decoherence 
rate are determined. In particular, we show that squeezing the bath to protect  
a non squeezed cat state against decoherence is equivalent to orthogonally squeezing 
the initial cat state while letting the bath be phase insensitive.
\end{abstract}
\section{Introduction}
In Schr\"odinger original paper \cite{schro}, a bipartite entangled state of the form
$$|\psi \rangle \propto |A\rangle |-\rangle + |B\rangle |+\rangle \: , $$ 
where $|A\rangle$ and $|B\rangle$ denote the ``alive'' and ``dead'' states of a cat
and $|\pm \rangle$ two orthogonal states of a microscopic system, was suggested
to illustrate the counterintuitive consequences of quantum mechanics in
a macroscopic setting.
More generally, in the literature, any single-system coherent superposition 
$|\psi \rangle \propto |\psi_{-} \rangle + |\psi_{+} \rangle$ of two pure 
quantum states $|\psi_{\pm} \rangle$ that are mesoscopically distinguishable, is 
often referred to as a {\em Schr\"odinger cat} or a {\em cat-like} state. 
The interest in the study of such superpositions, possibly
involving states of a microscopic system as well, stems from both theoretical 
and experimental considerations. 
Actually, a cat state is one of the simplest and most fundamental 
configurations allowing to probe the archetypal aspects of quantum theory: 
the superposition principle and quantum entanglement.

As far as quantum optical systems are concerned, the possibility of realizing 
superposition states of the radiation field, first envisioned by Yurke and Stoler 
\cite{yusto}, has been extensively investigated in later years 
\cite{theor,paris99}. However, pure state superpositions are in 
general corrupted by the interaction with 
the environment. Therefore, cat-like states that are available for experiments 
are usually mixed states that suffered a partial decoherence, and it is 
crucial to know whether and to which extent superpositions can survive the 
environmental noise. The theme of decoherence of cat--like states  
spurred relevant theoretical works \cite{walls85,kennedy,vitali,el-ora}, 
especially aimed to select schemes of quantum control and feedback 
stabilizing coherent superpositions against decoherence \cite{vitali}. 
Recent promising experimental results and perspectives continue to keep 
a widespread interest in this subject \cite{haroche,walther}.
\par
In the present paper, we study the non-unitary evolution of coherent and 
squeezed-coherent single-mode Schr\"odinger cat-like states in  
generic Gaussian noisy channels, namely, either 
thermal or squeezed-thermal baths of harmonic oscillators \cite{sqbath1,sqbath}. 
In particular, we will 
focus our attention on the evolution of the purity (or linear entropy) of the states, 
showing how the quantum superposition is corrupted by the interaction with a 
noisy environment and how to optimize the state and channel parameters
to minimize the decoherence rate.
\par
The paper is structured as follows. In Section \ref{s:cat} we introduce 
notations and evaluate the Wigner function of generalized cat-like states.
In Section \ref{s:evo} the time evolution in Gaussian noisy channels is
studied, whereas in Section \ref{s:zoo} the dependence of decoherence and
purity on the parameters of the initial state and of the channel is 
discussed in detail. In Section \ref{s:opt} optmized regimes
to minimize decoherence are discussed and, finally, Section \ref{s:out} 
provides some concluding remarks. 
\section{Cat-like states}\label{s:cat}
The simplest example of a Schr\"odinger cat state of a single-mode
radiation field is the following normalized superposition of coherent states 
\begin{equation}
|\alpha_{0},\theta\rangle \equiv 
\frac{|\alpha_{0}\rangle + \;{\rm e}^{i\theta}|-\alpha_{0}\rangle}
{\sqrt{2+2\cos(\theta)\;{\rm e}^{-2|\alpha_{0}|^{2}}}} \; .
\end{equation}
We denote by $\varrho_{\alpha_{0},\theta}$ the corresponding density matrix, 
whose symmetrically ordered characteristic function is given by 
\begin{eqnarray}
 \chi_{\alpha_{0},\theta} (\eta ) 
 \equiv  {\rm Tr}[\varrho_{\alpha_{0},\theta}D(\eta)]
& = &\frac{1}{\pi }\int \langle\alpha|\varrho_{\alpha_{0},\theta}
|\eta +\alpha \rangle\,\mathrm{e}^{(\eta \alpha ^{*}-\eta ^{*}\alpha )/2}\,\mathrm{%
d}^{2}\alpha \nonumber \\
&=&  \Big(2+2\cos(\theta)\;{\rm e}^{-2|\alpha_{0}|^{2}}\Big)^{-1}
\,{\rm e}^{-|\eta|^{2}/2} \nonumber \\
&&\times \Big[ 2\cosh(
\alpha_{0}^{*}\eta-\alpha_{0}\eta^{*}) +
\,{\rm e}^{-2|\alpha_{0}|^{2}}
2\cosh(\alpha_{0}^{*}\eta+\alpha_{0}\eta^{*}+i\theta)
\Big]\; ,
\end{eqnarray}
where  $D(\eta)=\exp(\eta  a^{\dag}-\bar\eta a)$ denotes
the displacement operator.
The corresponding Wigner function is defined as 
\begin{equation}
W_{\alpha_{0},\theta}(\alpha ) \equiv \frac{1}{\pi ^{2}}\int 
\mathrm{e}^{\eta ^{*}\alpha -\eta \alpha ^{*}}\chi_{\alpha_{0},\theta} (\eta )
\, \mathrm{d}^{2}\eta \; .
\label{cocat}
\end{equation}
From now on, let us move to quadrature variables $x$ and $p$, defined through 
$\alpha=(x+ip)/\sqrt{2}$. By defining
\begin{equation}
\boldsymbol{\widetilde \sigma}\equiv 
\left(\begin{array}{cc}
\frac{1}{2}&0\\
0&\frac{1}{2}
\end{array}\right) \; , \; \; \; \; X\equiv{x \choose p} \; ,
\end{equation}
one can write the Wigner function $W_{\alpha_{},\theta}(x,p)$ as follows
\begin{eqnarray}
 W_{\alpha _{0},\theta}(x,p) &=&\left(2\pi (1+\cos(\theta)\,\mathrm{e}%
^{-(x_{0}^{2}+p_{0}^{2})})\sqrt{{\rm Det}\,\boldsymbol{\widetilde \sigma}}\right)
^{-1}\nonumber\\
&&\times\left[ 
\,\mathrm{e}^{-\frac12(X^{T}-(x_{0},p_{0}))
\boldsymbol{\widetilde \sigma}^{-1}(X-%
{x_{0} \choose p_{0}})}\right.  
+\,\left.\mathrm{e}^{-\frac12(X^{T}+(x_{0},p_{0}))\boldsymbol{\widetilde \sigma}%
^{-1}(X+{x_{0} \choose p_{0}})}\right.\nonumber\\
&&+\left.\,\mathrm{e}^{-(x_{0}^{2}+p_{0}^{2})}\left( 
\,\mathrm{e}^{-\frac12(X^{T}-i(-p_{0},x_{0}))\boldsymbol{\widetilde \sigma}^{-1}(X-i%
{-p_{0} \choose x_{0}})+i\theta} + c. c.\right) \right] \; . 
\end{eqnarray}
The first two Gaussian terms are related to the projectors 
$|\alpha_{0}\rangle\langle\alpha_{0}|$ and 
$|-\alpha_{0}\rangle\langle-\alpha_{0}|$: they are the Wigner 
functions of the two coherent states $|\alpha_{0}\rangle$ and
$|-\alpha_{0}\rangle$. The remaining two terms correspond 
to non diagonal operators and are responsible for the intereference effects 
which characterize a coherent superposition.
\par
We now move to the study of a ``squeezed cat'', defined as 
the superposition of two squeezed coherent states. 
Let us introduce the operator $b$ by means of a Bogoliubov 
transformation
\begin{equation}
b\equiv\mu a+\nu a^{\dag}\; , \quad 
{\rm with} \quad|\mu|^{2}-|\nu|^{2}=1\; .
\label{eq:bogo}
\end{equation}
and the states $|\beta\rangle$ as its eigenvectors:
$b|\beta\rangle=\beta|\beta\rangle$. Such states are known
in the literature as `two--photon coherent states' and are indeed 
squeezed coherent states, according to the 
following well known relation \cite{walmil}
\begin{equation}
|\beta\rangle=D(\alpha)S(r_{0},\varphi_{0})|0\rangle\; , 
\end{equation}
with the requirements
\begin{equation} 
\alpha = \mu\beta-\nu\beta^{*}\; ,\quad
\cosh r_{0}=\mu \; ,  \quad
{\rm e}^{i2\varphi_{0}}\sinh r_{0}=\nu\; , \label{sqpar}
\end{equation}
and the squeezing operator defined as $S(r,\varphi)=\exp(
\frac12 r \,{\rm e}^{-i2\varphi} a^{2}-\frac12 r \,{\rm e}^{i2\varphi} a^{\dag 2})$. \par
We consider the following superposition
\begin{equation}
|\beta_{0},\theta\rangle \equiv 
\frac{|\beta_{0}\rangle + \;{\rm e}^{i\theta}|-\beta_{0}\rangle}
{\sqrt{2+2\cos(\theta)\;{\rm e}^{-2|\beta_{0}|^{2}}}}\: ,
\end{equation}
where the states $|\mp\beta_{0}\rangle$ are eigenstates of $b$: 
such a state will be referred to as to a ``squeezed cat'' state. \\
The Wigner representation of the state $|\mp\beta_{0}\rangle$ 
can be easily found by recalling that a two--photon coherent 
state $|\beta_{0}\rangle$ may as well be written as
\begin{eqnarray*}
|\beta_{0}\rangle=S(r_{0},\varphi_{0})D(\beta_{0})|0\rangle \; ,
\nonumber
\end{eqnarray*}
with the squeezing parameters $r_{0}$ and $\varphi_{0}$ determined by 
Eqs.~(\ref{sqpar}). This means that one can promptly derive the 
Wigner function $W_{\beta_{0},\theta}$ of a squeezed cat state by simply 
replacing $\alpha_{0}$ with $\beta_{0}$ in Eq.~(\ref{cocat}) and then 
applying a squeezing transformation.
In the following we will set $\varphi_{0}=0$, without loss of
generality, as a reference choice for phase space rotation.\\
The squeezing transformation implemented by $S(r_{0},0)$ 
corresponds, in terms of the phase--space variables 
$X={x \choose p}$ to the map
\begin{equation}
X\rightarrow{\bf R}^{-1}X \, , \quad {\rm with}\quad 
{\bf R}= \,{\rm diag}\,(\,{\rm e}^{r_{0}},\,{\rm e}^{-r_{0}})  \, .
\end{equation}
Applying such a transformation to the coherent cat Wigner function eventually
yields
\begin{eqnarray}
 W_{\beta _{0},\theta}(x,p)&=&\frac{1}{2\pi (1+\cos(\theta)\,\mathrm{e}%
^{-(x_{0}^{2}+p_{0}^{2})})\sqrt{{\rm Det}\,\boldsymbol{\sigma}_{0}}}
\nonumber\\
&&\times \left[ 
\,\mathrm{e}^{-\frac12(X^{T}-(x_{0},p_{0}){\bf R})
\boldsymbol{\sigma}_{0}^{-1}(X-%
{\bf R}{x_{0} \choose p_{0}})}\right.  
+\left.\,\mathrm{e}^{-\frac12(X^{T}+(x_{0},p_{0}){\bf R})\boldsymbol{\sigma}_{0}%
^{-1}(X+{\bf R}{x_{0} \choose p_{0}})}\right.\nonumber\\
&&+ \left.\,\mathrm{e}^{-(x_{0}^{2}+p_{0}^{2})}\left( 
\,\mathrm{e}^{-\frac12(X^{T}-i(-p_{0},x_{0}){\bf R})\boldsymbol{\sigma}_{0}^{-1}(X-i%
{\bf R}{-p_{0} \choose x_{0}})+i\theta}+c.c.\right) \right] \; ,
\end{eqnarray}
with 
\begin{equation}
\boldsymbol{\sigma}_{0}\equiv {\bf R}\boldsymbol{\widetilde \sigma}{\bf R}
\; .
\label{sqcat}
\end{equation}
Of course, for $r_{0}=0$ and $\beta_{0}=\alpha_{0}$, one recovers Eq.~(\ref{cocat}) 
for a coherent cat.  
\section{Time evolution in noisy channels}\label{s:evo}
We now consider the evolution in time of an initial squeezed cat state 
put in a noisy channel, in presence of damping and/or pumping
toward an asymptotic squeezed thermal state. 
The system is governed, in the interaction picture,
by the following master equation \cite{walmil}
\begin{equation}
{\dot\varrho  =}  \frac{\Gamma }{2}N \: L[a^{\dag}]\varrho
+\frac{\Gamma}{2}(N+1)\:L[a]\varrho -
\frac{\Gamma}{2}\: \Big( M^{*}\:D[a]\varrho + M
\:D[a^{\dag}]\varrho \Big)
\label{rhoev} \, ,
\end{equation}
where the dot stands
for time--derivative,
the Lindblad superoperators are defined by
$L[O]\varrho \equiv  2 O\varrho O^{\dag} -
O^{\dag} O\varrho -\varrho O^{\dag} O$ and
$D[O]\varrho \equiv  2 O\varrho O -O O\varrho -\varrho O O$,
and $M$ is the correlation function of the bath (which is 
usually referred  to as the squeezing of the bath).
It is in general a complex number $M \equiv M_1 + iM_2$, 
while $N$ is a phenomenological parameter related to
the purity of the asymptotic state. 
Positivity of the density matrix imposes
the constraint $|M|^{2} \le N(N+1)$.
At thermal equilibrium, {\it i.e.}~for $M=0$, $N$
coincides with the average number of thermal photons
in the bath. 
\par
As is well known, Eq.~(\ref{rhoev}) can be trasnsformed into 
a linear Fokker--Planck equation for the Wigner function of the system \cite{walmil}. 
Moreover, the Gaussian solutions of such equation have been 
thoroughly analysed in previous works \cite{kennedy,purezza}.
The initial condition we consider here is described by the Wigner function 
$W_{\beta_{0},\theta}$ of Eq.~(\ref{sqcat}), which is just a
linear combination of Gaussian terms. Therefore, its evolution in  
the noisy channel can be followed straightforwardly by exploiting the
general results derived in Ref.~\cite{purezza}: each Gaussian
term evolves independently and it suffices to follow the time dependence
of its first and second statistical moments.
\par
Let $\sigma_{ij}\equiv \frac12 \langle\hat x_{i}\hat x_{j}
+\hat x_{j}\hat x_{i}\rangle-\langle\hat x_{i}\rangle\langle\hat x_{j}\rangle$ 
and $X_{0i}\equiv\langle\hat x_{i}\rangle$ be, respectively, the 
covariance matrix and the vector of the first moments of a Gaussian state
($\hat x_{1},\hat x_{2}=
\hat x,\hat p$ being the quadrature phase operators).
Then, the time--evolution of $\boldsymbol{\sigma}(t)$ and $X_{0}(t)$ 
in the squeezed thermal channel is
described by the following Eqs.~\cite{purezza}
\begin{eqnarray}
X_{0}(t)  =  \,{\rm e}^{-\frac\Gamma2 t}X_{0}(0) \; , \\
\nonumber\\
\boldsymbol{\sigma}(t)  =  \boldsymbol{\sigma}_{\infty}
\left( 1-\,{\rm e}^{-\Gamma t} \right) + \boldsymbol{\sigma}(0)
\,{\rm e}^{-\Gamma t} \; , \label{sigma}\\
\nonumber\\
{{\rm with}\quad}\boldsymbol{\sigma}_{\infty}=\left(\begin{array}{cc}
\frac{(2N+1)+2M_{1}}{2}&M_{2}\\
&\\
M_{2}&\frac{(2N+1)-2M_{1}}{2}\end{array}\right)  . \label{sigmainf}
\end{eqnarray}
The time-dependent solution for the Wigner function $W_{\beta_{0},\theta}(t)$ 
of an initial squeezed cat is thus readily found and reads
\begin{eqnarray}
 W_{\beta _{0},\theta}(x,p) & = & \left(2\pi (1+\cos(\theta)\,\mathrm{e}%
^{-(x_{0}^{2}+p_{0}^{2})})\sqrt{{\rm Det}\,\boldsymbol{\sigma}(t)}
\right)^{-1}
\nonumber\\
&&\times
\left[ 
\,\mathrm{e}^{-\frac12(X^{T}-\,{\rm e}^{-\frac\Gamma2 t}(x_{0},p_{0}){\bf R})
\boldsymbol{\sigma}(t)^{-1}(X-%
\,{\rm e}^{-\frac\Gamma2 t}{\bf R}{x_{0} \choose p_{0}})}\right.  \nonumber\\
&&+\,\mathrm{e}^{-(x_{0}^{2}+p_{0}^{2})}\left( 
\,\mathrm{e}^{-\frac12(X^{T}-i\,{\rm e}^{-\frac\Gamma2 t}(-p_{0},x_{0}){\bf R})
\boldsymbol{\sigma}(t)^{-1}(X-i%
\,{\rm e}^{-\frac\Gamma2 t}{\bf R}{-p_{0} \choose x_{0}})+i\theta} + c. c.\right)
\nonumber\\
&&+\left. \mathrm{e}^{-\frac12(X^{T}+\,{\rm e}^{-\frac\Gamma2 t}(x_{0},p_{0})
{\bf R})\boldsymbol{\sigma}(t)%
^{-1}(X+\,{\rm e}^{-\frac\Gamma2 t}{\bf R}{x_{0} \choose p_{0}})}
\right] \; ,
\label{wigner}
\end{eqnarray}
with $\boldsymbol{\sigma}(t)$ given by Eqs.~(\ref{sigma}, \ref{sigmainf}). 
The first moments of each Gaussian term are exponentially damped in the channel. 
Any initial cat state is
attracted toward an asymptotic centered squeezed thermal state with Wigner function 
\begin{equation}
W_{\infty}(x,p)=
\frac{{\rm e}^{-\frac12 X^{T}\boldsymbol{\sigma}_{\infty}^{-1}X}}
{\pi \sqrt{{\rm Det}\,\boldsymbol{\sigma}_{\infty}}} \; .
\label{asy}
\end{equation}
This state, like all asymptotic quantities, is a property of the channel and 
does not depend on the initial state.
\section{Decoherence of an initial cat state}\label{s:zoo}
In order to quantify  
decoherence of the state caused by environmental noise, we consider 
the loss of purity. The degree of purity of a continuous variable quantum
state $\varrho$ can be effectively characterized either by its Von Neumann entropy 
$S_{V}\equiv-\,{\rm Tr}\,(\varrho\ln\varrho)$ or by its linear entropy $S_{l}$
\begin{equation}
S_{l}  \equiv  1-\,{\rm Tr}(\varrho^{2})  \equiv 1-\mu
=1-\frac\pi2
\int_{\mathbb{R}^2} W^{2}\,{\rm d}x\,{\rm d}p \: .\label{purea}
\end{equation}
In the following we will adopt linear entropy, which can be conveniently
evaluated.
The quantity $\mu = {\rm Tr}\varrho^{2}$, conjugate to $S_{l}$, will be 
referred to as the
`purity' from now on.\par
For an initial pure cat state ($\mu=1$) at time $t=0$, the  
asymptotic value $\mu_{\infty}$ can be obtained from Eq.~(\ref{asy})
by straightforward integration \cite{purezza}
\begin{equation}
\mu_{\infty}=\frac{1}{2\sqrt{\,{\rm Det}\,\boldsymbol{\sigma}_{\infty}}}
=\frac{1}{\sqrt{(2N+1)^{2}-4|M|^{2}}} \, .
\end{equation}
At finite times the purity of a decohering cat
can be determined integrating the function $W_{\beta
_{0},\theta}(x,p)$ given by Eq.~(\ref{wigner}), according to
Eq.~(\ref{purea}). The integration can be promptly 
performed with the help of 
the following thumb rule
\begin{eqnarray*}
{{\rm if}\quad X_{1}}\equiv\left(\begin{array}{c}x-x_{1}\\p-p_{1}\end{array}\right)&,\quad
X_{2}\equiv\left(\begin{array}{c}x-x_{2}\\p-p_{2}\end{array}\right)&,{\rm and}\quad
\bar X\equiv\left(\begin{array}{c}x_{1}-x_{2}\\p_{1}-p_{2}\end{array}\right) \; ,
\end{eqnarray*}
\begin{equation}
{{\rm then}}\quad\int{\rm e}^{-\frac12 X_{1}^{T}\boldsymbol{\sigma}^{-1}X_{1}}
{\rm e}^{-\frac12 X_{2}^{T}\boldsymbol{\sigma}^{-1}X_{2}}\,{\rm d}x\,{\rm d}p=
\pi\sqrt{{\rm Det}\,\boldsymbol{\sigma}}\;{\rm e}^{-\frac{1}{4}\bar X^{T}\boldsymbol{\sigma}^{-1}
\bar X}\, .
\end{equation}
Exploiting the above rule one eventually has
\begin{eqnarray}
 \mu_{\beta_{0},\theta}(t) &=& \left(8(1+\cos(\theta)\,{\rm e}^{-(x_{0}^{2}+p_{0}^{2})})^{2}
\sqrt{{\rm Det}\,\boldsymbol{\sigma}(t)}\right)^{-1}\nonumber\\
&&{\times \Bigg[2}\left(1+{\rm e}^{-\,{\rm e}^{-\Gamma t}X_{0}^{T}
{\bf S}(t)X_{0}}\right)+
2\,{\rm e}^{-2(x_{0}^{2}+p_{0}^{2})}\left(\cos(2\theta)+{\rm e}^{\,{\rm e}^{-\Gamma t}
X_{0}^{T}
{\bf T}(t)X_{0}}\right)\nonumber\\
&&{+ 
4\,{\rm e}}^{-(x_{0}^{2}+p_{0}^{2})}\cos(\theta)
\left( {\rm e}^{-\,{\rm e}^{-\Gamma t}(x_{0}+ip_{0})^{2}
A(t)} + c. c.\right)\Bigg] \, ,
\label{purezza}
\end{eqnarray}
with
\begin{equation}
A(t)\equiv \frac{S_{xx}(t)-S_{pp}(t)-2iS_{xp}(t)}{4} 
\end{equation}
and
\begin{equation}
{\bf S}(t) \equiv {\bf R}\boldsymbol{\sigma}(t)^{-1}{\bf R} \; , \quad
{\bf T}(t)\equiv ({\rm Det}\,\gr{\sigma})^{-1}{\bf S}(t)^{-1}
\end{equation}
Eq.~(\ref{purezza}) shows that $\mu_{\beta_{0},\theta}$ is a decreasing function of
the initial parameters $x_{0}$ and $p_{0}$. This should be expected: 
the `bigger' the cat is, the faster it decoheres. For $\beta_{0}=0$ the 
superposition disappears and the cat state 
reduces to a centered squeezed state; the latter, whose evolution in noisy channels
has been studied in Ref.~\cite{purezza}, decoheres more slowly 
than the corresponding cat state.
\begin{figure}[t!]
\begin{center}
\epsfbox{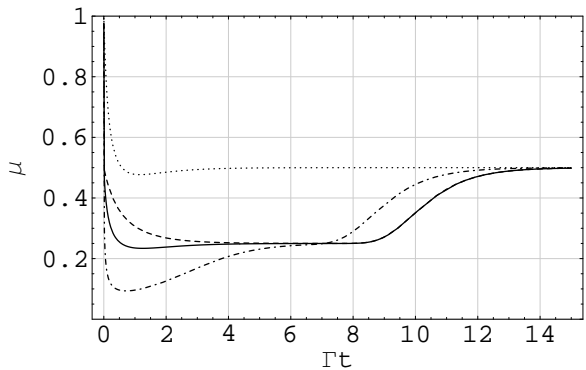}
\caption{Evolution of the purity for different initial cat states and channels. 
In all cases the asymptotic purity of the bath is $\mu_{\infty}=0.5$.
The dotted line shows the behaviour of an initial cat 
with $x_{0}=p_{0}=1$ and $r_{0}=0$ in a non squeezed channel. 
The dashed and the continuous lines refer to an initial cat with $x_{0}=p_{0}=100$ 
and $r_{0}=0$ evolving, respectively, in a non 
squeezed channel and in a squeezed channel with $M=2+2i$. The dot--dashed line refers 
to an initial cat state with $x_{0}=p_{0}=10$ and $r_{0}=2$, evolving in
a non squeezed channel.}
\label{f:cats}
\end{center}
\end{figure}
The numerical analysis shows that
the phase $\theta$ has little effect on the behaviour of purity at large times: 
in fact, as it is evident from Eq.~(\ref{purezza}), all the terms involving  $\theta$
are suppressed by Gaussian terms of the form ${\rm Exp}\,(-x_{0}^2-p_{0}^2)$.
In Fig.~\ref{f:cats} the behaviour of purity over the full temporal range up to 
the asymptotic regime is showed for various choices of the parameters of the channel
and of the initial state. \par
One feature which is most evident in all instances is the fast initial fall of the purity. 
Although the minimum value of the purity attained in such a steep descent can vary, 
the temporal scale in which the minimum is reached is always
of the order of $\Gamma^{-1}$, which is indeed the only time characterizing 
the losses in the channel. 
Besides, it can be seen that the general  
behaviour of purity in squeezed baths is quite the same as in 
non squeezed ones. One can see as well that the value of the first minimum of the
purity depends drastically on the squeezing parameter $r_{0}$, decreasing with greater $r_{0}$,
while, for a given squeezing parameter, an increase in the parameters 
$x_{0}$ and $p_{0}$ delays the reaching of the asymptotic purity (see Fig.~\ref{f:cats}).

\section{Optimal regimes}\label{s:opt}

Optimal regimes with minimized decoherence can be determined 
by maximizing the purity at any given time for fixed values of the parameters
of the channel and of the initial state.  
Notice that, as we have shown in the previous section, a cat--like 
state decoheres on a time scale of the order of the photon lifetime 
$\Gamma^{-1}$, regardless of the choice of the parameters of both 
the Gaussian reservoir and the initial pure cat state. This fact is a manifestation 
of a fundamental feature of quantum mechanics: once a single photon 
is added or lost, all the information contained in a coherent superposition
`leaks out to the environment' and 
is therefore lost as well, together with the possibility of detecting 
such a coherent behaviour by means of interferometry \cite{lee,harochelh}.
A simple, meaningful example in this respect is just a coherent 
even cat $\ket{\alpha_{0},0}$ subjected to damping. 
Under the loss of a photon, such a state 
jumps into $a\ket{\alpha_{0},0}\propto\ket{\alpha_{0},\pi}$, which 
is an odd cat and has got opposite interference terms. Therefore, as soon 
as the probability of losing a photon reaches $0.5$, the original 
superposition turns into an incoherent mixture of an even and an odd
cat, whose interference terms cancel out each other exactly \cite{harochelh}.\par 
Actually, a more detailed analysis would reveal that decoherence times 
are even shorter than $\Gamma^{-1}$: for a coherent cat $\ket{\alpha_{0},\theta}$
evolving in a thermal environment, 
coherence is lost at $t_{dec}=\Gamma^{-1}/2|\alpha_{0}|^2$ \cite{walls85}.
In view of these considerations, we
are interested in maximizing the purity in the time region 
$\Gamma t\lesssim 1/2|\beta_{0}|^2$.\par 
A relevant question in such a context is: given an initial bath and an 
initial squeezing of the cat--like state,
which is the optimal phase space direction 
of $\beta_{0}\equiv|\beta_{0}|\,{\rm e}^{i\xi}$ 
at fixed $|\beta_{0}|$? Note that the last condition 
can be seen as a constraint on the energy of the cat--like state. 
Obviously, for a non squeezed cat in a thermal channel the 
symmetry of the problem forbids the existence of a privileged direction.\par 
Interesting issues come instead from the consideration of a squeezed 
cat in a thermal channel and a non squeezed cat in a squeezed bath.
For the moment being, let us consider the instance of an `even' cat, {\it i.e.}~of a cat 
with coherent phase $\theta=0$.
The dependence of the purity on $\beta_{0}$ is essentially contained in  
exponentials of quadratic forms, see Eq.~(\ref{purezza}). 
The algebraic analysis of such terms 
in the case of a squeezed cat in thermal baths and of a coherent cat in 
squeezed baths is quite easy (see App.~\ref{app}). 
If the squeezing is performed on the inital cat state, 
the coherence is 
better preserved if $\beta_{0}$ is chosen in the same phase space direction 
of the variance which is suppressed by squeezing. With our notation, this corresponds
to $\xi=\varphi_{0}+\pi/2$ (where $\varphi_{0}$ is the squeezing angle).
On the contrary, if the squeezing is performed on the bath in the direction 
$\varphi_{\infty}$, the choice $\xi=\varphi_{\infty}$ turns out to be the best one to 
slow down the rate of decoherence.
This somehow counterintuitive situation is due to the existence 
of complex fringe patterns of a cat state in phase space, especially in 
the presence of squeezing. Actually, preserving 
quantum coherence is crucially related to the persistance
of the interference fringes: reducing quantum  
fluctuations in a phase space direction protects the fringes from degradation and 
the cat state from decoherence. 
With the above expedient choices of the phase $\xi$, 
squeezing does actually improves the coherence of the superposition at small times
with respect to dissipation in a phase insensitive setting \cite{kennedy}.\par
Now, let us suppose that for an even cat state the phase $\xi$ is optimally chosen and 
let us consider a channel with asymptotic purity $\mu_{\infty}$.
It can be easily shown (cfr.~App.~\ref{app}) that 
the purity of an initially squeezed cat (with squeezing $r$) in such a thermal 
channel equals, at any given time, the one of a non squeezed cat evolving 
in a squeezed channel with the same squeezing $r$. The same protection 
against decoherence provided by the squeezing of the bath can be obtained, in a thermal
phase insensitive channel, by squeezing the initial even cat state of the same amount 
in an orthogonal direction.\par
\begin{figure}[t!]
\begin{center}
\epsfbox{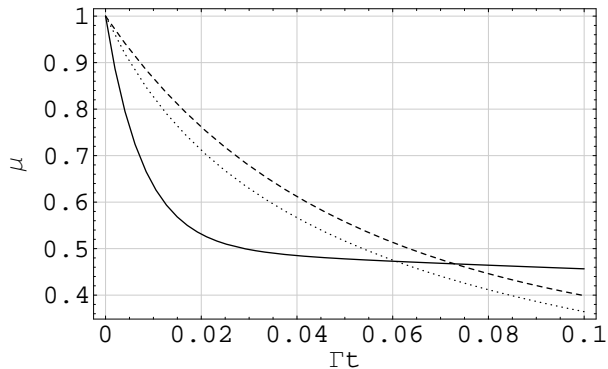}
\caption{Comparison between the evolution of purity
of an initial non squeezed cat in a non squeezed bath (continuous line) and of
initial cats in squeezed configurations with optimal phase choices. 
The dashed line refers to an initial squeezed cat with $r_{0}=1$, whereas 
the dotted line refers to an initial squeezed cat with $r_{0}=1.5$. 
In all instances $\mu_{\infty}=0.5$, $|\beta_{0}|^2 =16$ and $\theta=0$.
The decoherence time for the non squeezed cat is $t_{dec}\simeq0.03\Gamma$,
in good agreement with the initial decrease of purity. 
The choice $r_{0}\simeq1$ appears to be optimal for such 
a value of $|\beta_{0}|$.}
\label{smtim}
\end{center}
\end{figure}

We finally remark that, with an optimal phase setting, an optimal finite value
of the squeezing parameter $r$ does indeed exist (see Fig.~\ref{smtim}). 
This fact can be best appreciated from Eq.~(\ref{purezza}):
even if the exponential terms increase with increasing $r$, the factor ${\rm Det}\,
\gr{\sigma}^{-1/2}$ decreases with $r$. The value of $r$ allowing 
the maximum slowing down of decoherence increases with increasing $|\beta_{0}|$.\par
We now briefly consider the instance $\theta=\pi/2$, which is the one 
considered originally by Yurke and Stoler as being produced 
by a unitary evolution in Kerr media. In such a case, the last term 
in Eq.~(\ref{purezza}) vanishes, so that the optimal choice for $\xi$ can be 
easily found for any choice of $r_{0}$, $r_{\infty}$ and $\varphi_{\infty}$.
To this end, it is sufficient to determine the eigenvector corresponding to the smallest eigenvalue 
of the matrix ${\bf S}(t)^{-1}$, which coincides with the analogous eigenvector 
of $\gr{R}^{-1}\gr{\sigma}_{\infty}\gr{R}^{-1}$ and does not depend on time. 
However, we once again stress that, as soon as one deals with mesoscopic cat states (so that  
${\rm Exp}\,({-|\beta_{0}|^2})\ll 1$), the dependence on the coherent phase 
$\theta$ is severely suppressed and all the considerations made for even cat
states still hold, regardless of the choice of $\theta$.
Results for the evolution of purity at small times are summarized in Fig.~\ref{smtim}.
\section{Conclusions}\label{s:out}
The study of the decoherence of initial coherent and squeezed coherent
cat states in arbitrary Gaussian reservoirs has been 
carefully carried out by determining the exact time evolution of the purity of the state. 
Optimal settings that minimize the rate of decoherence in relevant configurations 
have been determined.\par
In particular, we have shown that the same protection against decoherence
granted by a squeezed bath can be achieved by squeezing the initial 
cat--like state. In view of the well known difficulties involved in the experimental
realization of squeezed baths, even as effective descriptions of feedback schemes 
\cite{vitali}, this equivalence could provide a relevant alternative
option for experimental purposes \cite{paris99}.

\subsection*{Acknowledgements}
We thank David Vitali for useful discussions. AS, FI and SDS 
thank INFM, INFN, and MIUR under national project PRIN-COFIN 2002 
for financial support. MGAP is a research fellow at {\em Collegio Volta},
and he thanks R. F. Antoni for continuing inspiration.

\appendix
\section{Appendix \label{app}}
In this appendix we analytically single out the optimal phase space 
orientations of the cat state for the configurations discussed in Section~\ref{s:opt}. 
The orientation of the cat state is determined by the angle $\xi=\,{\rm Arg}(\beta_{0})$.
In the following we will set $\theta=0$ and define, for ease of notation, $u\equiv
(1-\,{\rm Exp}(-\Gamma t))/2\mu_{\infty}$ and $v\equiv\,{\rm Exp}(-\Gamma t)/2$. 
The squeezing parameter $r_{\infty}$ of the bath is determined 
by $\cosh(2r_{\infty})=\sqrt{1+4\mu_{\infty}^{2}|M|^{2}}$ \cite{purezza}.

We first consider a squeezed cat in a thermal 
channel, with $r_{0}\neq0=r_{\infty}$. In this instance, one has
$${\bf S}(t)=\,{\rm diag}\{\,{\rm e}^{2r_{0}}(u+\,{\rm e}^{2r_{0}}v)^{-1},
\,{\rm e}^{-2r_{0}}(u+\,{\rm e}^{-2r_{0}}v)^{-1}\} \; ,$$
and 
$$A(t)=(2\,{\rm Det}\,\gr{\sigma}(t))^{-1}u\sinh(2r_{0}) \; .$$
Substituting these expressions in Eq.~(\ref{purezza}), 
it is easy to see that, for fixed $2|\beta_{0}|^{2}=
x_{0}^{2}+p_{0}^2$,
all the exponential terms are 
maximized by the choice $x_{0}=0$, $p_{0}=\sqrt{2}|\beta_{0}|$. This corresponds 
to $\xi=\pi/2=\varphi_{0}+\pi/2$.\par
Analogously, for a non squeezed cat in a squeezed channel (with $r_{0}=0\neq r_{\infty}$), 
one gets \footnote{We can set the squeezing angle of the bath $\varphi_{\infty}=0$ 
because the reference choice is still
free, being the initial cat non squeezed. This corresponds to choosing the squeezing $M$ real.}
$${\bf S}(t)=\,{\rm diag}\{\,{\rm e}^{-2r_{\infty}}(u+\,{\rm e}^{-2r_{\infty}}v)^{-1},
\,{\rm e}^{2r_{\infty}}(u+\,{\rm e}^{2r_{\infty}}v)^{-1}\} \; ,$$
and 
$$A(t)=-(2\,{\rm Det}\,\gr{\sigma}(t))^{-1}u\sinh(2r_{\infty}) \; .$$
Eq.~(\ref{purezza}) shows that 
the choice $x_{0}=\sqrt{2}|\beta_{0}|$, $p_{0}=0$ 
(corresponding to $\xi=0=\varphi_{\infty}$) 
is optimal in this case.\par
Finally, it is easy to verify that, 
adopting such optimal choices and putting $r_{0}=r_{\infty}$, 
all the exponential terms entering in Eq.~(\ref{purezza}) take 
the same values for a squeezed cat in a thermal channel and a non squeezed
cat in a squeezed channel. Since $\,{\rm Det}\,\gr{\sigma}$ depends only
on the difference $|r_{0}-r_{\infty}|$ \cite{purezza}, this implies that the time 
evolutions of purity are identical in these two instances.  

\end{document}